# Magnetoelectric exceptional points for isolated dielectric nanoparticles

Adrià Canós Valero[1*], Vjaceslavs Bobrovs[2], Dmitrii Redka[2], Thomas Weiss[1], Alexander S. Shalin[3] and Yuri Kivshar[4*]

[1]Institute of Physics, University of Graz, and NAWI Graz, Graz 8010, Austria

[2]Riga Technical University, Institute of Telecommunications, Riga 1048, Latvia

[3]Center for Photonics and 2D Materials, Moscow Institute of Physics and Technology, Dolgoprudny 141700, Russia

[4]Nonlinear Physics Centre, Department of Fundamental and Theoretical Physics, Australian National University, Canberra ACT 2601, Australia

**Abstract**

We study scattering of electromagnetic waves by nonspherical dielectric resonators and observe the appearance of the exceptional points in the eigenvalue spectrum being the key features of non-Hermitian systems. We demonstrate how breaking of the shape symmetry of an isolated dielectric particle is linked to the existence of exceptional points and formulate the general conditions for strong coupling of resonances. We discuss in detail the example of an electric and magnetic dipole modes supported by a silicon nanoparticle. We argue that any two modes of a resonant dielectric nanoparticle can merge to create an exceptional point provided their resonant frequencies cross as functions of particle's parameter such as e.g., aspect ratio, and their field distributions have opposite signs after a reflection in the transverse plane of the structure. The strongly coupled modes radiate as a mixture of electric and magnetic dipoles resulting in a strong magnetoelectric response, being easily controlled by symmetry-breaking perturbations. We also study the effect of a dielectric substrate and demonstrate that the substrate provides an additional mechanism to tune the position of exceptional points in the parameter space. Finally, we discuss applications of magnetoelectric exceptional points, including their use for refractive-index sensing.

**Introduction**

Open physical systems, such as systems interacting with an environment, are ubiquitous in nature. One of their most remarkable features is their ability to host spectral singularities where both eigenvalues and eigenvectors coalesce, known as *exceptional points* (EPs). At an EP, there is a collapse in the dimensionality of the eigenspectrum, resulting in a drastic increase in sensitivity to a small perturbation $\epsilon$. Unlike Hermitian degeneracies (such as diabolic points) whose eigenfrequencies shift linearly with a perturbation, and for n-th-order EP the shift is proportional to $\epsilon^{1/n}$, where $n$ is the number of coalescing eigenmodes. This peculiarity renders EPs highly attractive for many applications including biosensing schemes relying on resonance tracking[1].

EPs have no analogue in Hermitian systems, such as those employed in conventional quantum mechanics. However, other areas of wave physics are starting to explore the opportunities offered by operating at the vicinity of an EP. Besides their unique sensitivity, EPs are associated with abrupt phase transitions. In optics, this feature leads to a plethora of counterintuitive phenomena, such as unidirectional invisibility, enhanced Sagnac effect in microcavities[2], laser mode selectivity, loss-induced revival of lasing, bulk Fermi arcs[3] and topologically-protected chirality[4]. In the momentum space, they are associated with a half-integer polarization charge [5].

EPs are closely linked to Parity-Time (PT) symmetry, a notion inherited from quantum mechanics that tells us that systems invariant to parity inversion and time reversal can exhibit real spectra[6]. Indeed, it has been shown that EPs mark the transition between the PT-symmetric and PT-broken phase[7]. Most experimental realizations of EPs in photonics rely on this idea. Namely, by balancing gain and dissipation in optical microcavities, an EP emerges, drastically altering the response of the system[8,9].

Nevertheless, the presence of gain and loss or PT-symmetry is not necessary in order to reach an EP: they were recently demonstrated in plasmonic metasurfaces by balancing the losses of two plasmon modes[1,10]. Plasmonic EPs were utilized to develop an ultrasensitive scheme for anti-immunoglobulin G[1]. So far, the vast majority of works are restricted to wavelength-scale systems based on waveguides, metasurfaces and/or microresonators [1,3,4,9,11–13].

In recent years, resonant all-dielectric nanostructures have progressively replaced plasmonics in a broad range of applications in nano-optics[14,15]. This trend has been largely motivated by their negligible ohmic losses and their compatibility with Complementary Metal Oxide Semiconductor (CMOS) technology. Whereas plasmonic structures support plasmons, localized free electron oscillations at the surface of metallic particles, dielectric nanoparticles exhibit geometric resonances, concentrated within their volume. They are associated with the excitation of complex distributions of polarization currents[7]. Unlike conventional plasmonic particles[16], dielectric nanoparticles support a broad class of multipolar resonances of both electric and magnetic type, which can mutually interfere leading to remarkable phenomena such as the Kerker effect[17–21], nonradiating anapoles[22,23], spin to orbital angular momentum conversion[24], or magnetic light[25,26].

Intriguingly, even in the absence of heat dissipation, the modes of all-dielectric nanostructures are non-Hermitian in nature, as a result of radiation losses. Leveraging non-Hermiticity in dielectric nanostructures has recently enabled the realization of quasi bound states in the continuum in isolated cavities [27,28], and revealed a new superscattering regime accessible through modal interference [29].

Here, we aim at introducing non-Hermitian singularities as a new prospective tool for dielectric resonant nanophotonics. Specifically, we study the conditions necessary for their realization and observation in a single dielectric nanoparticle with simple topology. Critically, unlike the only early attempt[30], our novel design strategy is based solely on geometrical symmetry breaking[31,32], without changing the refractive index of the structure. This makes the EP topologically robust and realizable in a broad class of dielectric materials, including Si. While we illustrate our approach with the lowest order electric and magnetic dipole modes (ED and MD modes) of a Si nanodisk, the theory is general, allowing to design EPs from any two modes of a dielectric nanodisk fulfilling two conditions:

1. their resonant frequencies must cross as a function of aspect ratio,
2. their field distributions must have opposite mirror symmetry in the transverse plane of the structure.

In our example, we show that the EP corresponds to the onset of the strong coupling regime between the ED and MD modes. The modes radiate as a mixture of electric and magnetic dipoles, which result in a strong magnetoelectric response, easily controlled by the symmetry breaking perturbation (Figure 1a-b). The latter can be probed experimentally, for instance, by measuring the asymmetry in backscattering from illuminations from top and bottom, as we calculate in Figure 1a-b.

Furthermore, we investigate the influence of a dielectric substrate, demonstrating how the latter provides an additional mechanism to tune the EP position in parameter space, facilitating the experimental implementation. Finally, we showcase the potential of magnetoelectric EPs to detect small variations in the refractive index of the environment. The results confirm a much stronger sensitivity to perturbations in comparison with conventional ED and MD modes.

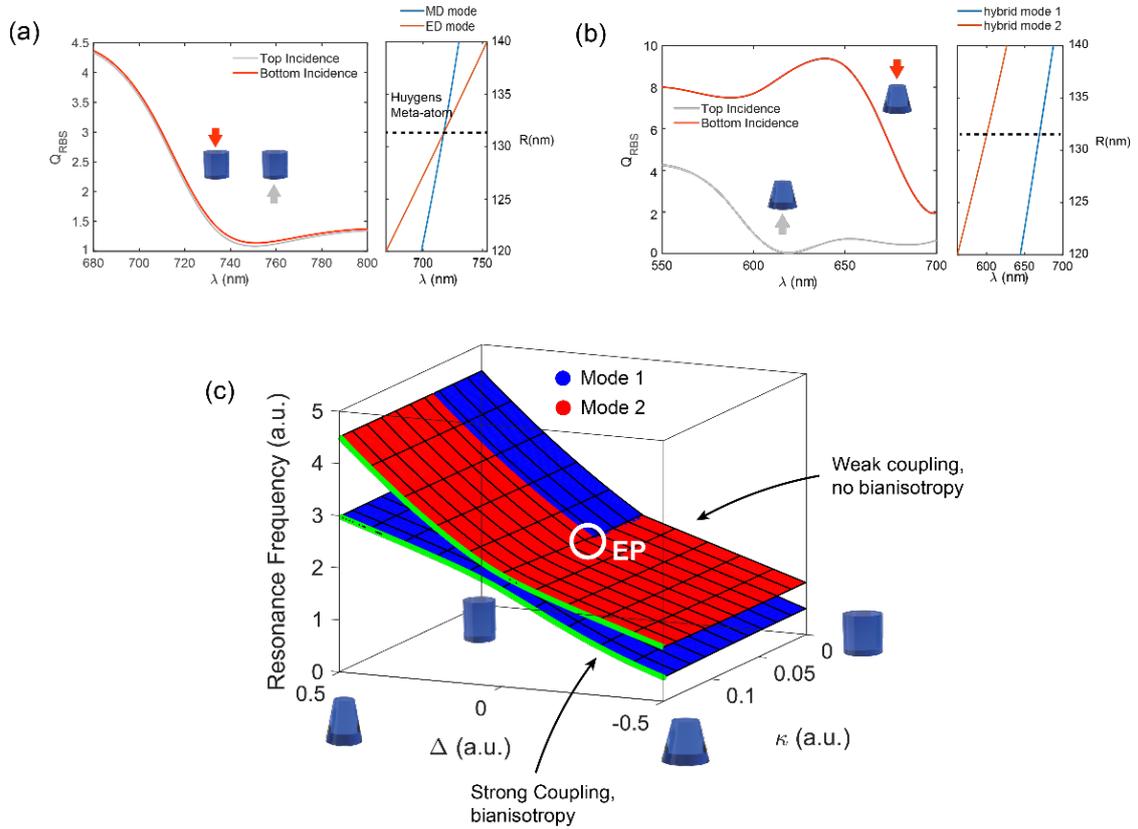

**Figure 1. Non-Hermitian topological transition from weak to strong bianisotropic response.** (a) Right panel: Radar backscattering cross section ($Q_{RBS}$), for a weakly truncated Si nanocylinder under normally incident plane wave illumination from top (red arrow in the left inset) and bottom (right inset). The top radius R has been optimized to induce a crossing between the electric and magnetic dipole modes (ED and MD modes), forming a Huygens meta-atom, as shown in the left panel. The black dashed line indicates the chosen radius. (b) Same as (a), for a truncated nanocone with top radius 80 nm. The bottom radius is fixed to the one in (a). Due to the breaking of the vertical mirror symmetry, the modes are now coupled and possess a mixed magnetoelectric character (hybrid modes 1 and 2). As a result, the back-scattering from the particle can be resonantly suppressed (enhanced) for illumination from the top (bottom). The decrease in the overall volume of the nanoparticle causes the resonances to blue-shift. Parameters of the nanoparticles: height 100 nm, top radius 131 nm. (c) Evolution of the resonant frequencies of a non-

Hermitian Hamiltonian, calculated with Eq.(5). Here, $\Delta$ denotes the detuning between the uncoupled modes and $\kappa$ is the coupling coefficient, assumed to be real (see details in text). An EP marks the transition from weak to strong coupling regimes.

**Two-mode model of a dielectric nanoparticle**

Consider a dielectric nanoparticle illuminated by a normal-incident plane wave. The plane wave excites several resonant modes in the particle. They can be understood as geometrical resonances, arising from the interference of traveling waves due to reflections from the cavity walls. They correspond to 'quasinormal' modes, solutions of Maxwell's equations with open boundary conditions in the absence of an exciting source [33,34].

For our purposes, we consider there are only two relevant modes in the spectral range of interest. We will denote them as $|m\rangle$ (as for magnetic dipole), and , $|p\rangle$ (as for electric dipole).

Suppose that we perform a small perturbation in the shape of such particle. Then, the modes of the perturbed particle, $|u,d\rangle$, can be expressed as a linear combination of the original ones, e.g. $|u\rangle = C_m|m\rangle + C_p|p\rangle$. This yields a generalized eigenvalue problem[35–37] for the coefficients $C_\alpha$:

$$\begin{pmatrix} \tilde{\omega}_p & 0 \\ 0 & \tilde{\omega}_m \end{pmatrix} \begin{pmatrix} C_p \\ C_m \end{pmatrix} = \tilde{\omega}_{u,d} \begin{pmatrix} 1+V_{pp} & V_{pm} \\ V_{mp} & 1+V_{mm} \end{pmatrix} \begin{pmatrix} C_p \\ C_m \end{pmatrix}. \quad (1)$$

Note that, since the modes leak energy to the environment, the eigenfrequencies $\tilde{\omega}_\alpha$ are complex, so that $\tilde{\omega}_\alpha = \omega_\alpha - i\gamma_\alpha$. $\omega_\alpha$ is the resonant frequency of the mode, while $\gamma_\alpha$ is the loss rate. For small $\gamma_\alpha$, the loss rate is proportional to the width of the resonance peak.

The $V_{\alpha\beta}$ elements in Eq.(1) represent the perturbation. They are rigorously calculated as integrals of the modal fields over the surface of the nanoparticle[36]:

$$V_{\alpha\beta} = \Delta\varepsilon \int_S \mathbf{E}_\alpha^{(+)}(\mathbf{r}) \cdot \Delta V(\mathbf{r}) \mathbf{E}_\beta^{(-)}(\mathbf{r}) d^2\mathbf{r} = \langle \mathbf{E}_\alpha \, \Delta V \, \mathbf{E}_\beta \rangle. \quad (2)$$

The (+,-) signs in the superscripts indicate that the field must be evaluated directly outside or inside the surface of the particle, respectively. $\Delta\varepsilon$ is the permittivity contrast between the particle and the environment. The shape perturbation is contained in $\Delta V(\mathbf{r})$, which is the deformation undergone by the particle normal to its boundary.

**Existence of exceptional points for a single dielectric nanoparticle**

We are now exploring the possibility to induce an EP in this system. First, we consider a simplified analytical form of Eq.(1), valid for modes with small loss rates:

$$\mathcal{H}_0 \begin{pmatrix} C_p \\ C_m \end{pmatrix} = \tilde{\omega}_{u,d} \begin{pmatrix} C_p \\ C_m \end{pmatrix}, \tag{3}$$

$$\mathcal{H}_0 = \begin{pmatrix} \tilde{\omega}_p^{(1)} & \kappa_{pm} \\ \kappa_{mp} & \tilde{\omega}_m^{(1)} \end{pmatrix} \tag{4}$$

where $\tilde{\omega}_\alpha^{(1)} = \tilde{\omega}_\alpha / (1 + V_{\alpha\alpha})$ and the 'coupling' coefficients $\kappa_{\alpha\beta} \propto V_{\alpha\beta}$. $\mathcal{H}_0$ can now be interpreted as an 'effective Hamiltonian' dictating the evolution of the modes. The perturbed eigenfrequencies are found as:

$$\tilde{\omega}_{u,d} = E_{pm} \pm \sqrt{\Delta_{pm}^2 + \kappa_{pm}\kappa_{mp}}, \tag{5}$$

$E_{pm}$ is the average of the two unperturbed eigenfrequencies, while $\Delta_{pm}$ is half their detuning, namely $\Delta_{pm} = (\tilde{\omega}_p - \tilde{\omega}_m)/2$.

Importantly, both real and imaginary parts of the perturbed eigenfrequencies become identical when the square root in Eq.(5) vanishes. This situation fully corresponds to an EP. Assuming the coupling takes place mainly through evanescent fields, the $\kappa_{\alpha\beta}$ coefficients become real. Furthermore, if the leakage to the environment is small, $\kappa_{pm} \approx \kappa_{mp}^*$. Two conditions are necessary to achieve the EP: (i) The resonant frequencies of the original modes must coincide so that $\omega_p = \omega_m$, and (ii) $2|\kappa_{pm}| \approx \pm(\gamma_p - \gamma_m)$. In conclusion, unlike conventional PT-symmetric structures, designing an EP in a passive, all-dielectric nanoparticle should require the ability to tune at least two parameters instead of one, in order to fulfill the two conditions above.

A prototypical example of the eigenfrequency evolution near a passive EP is displayed in Figure 1c. The latter has been calculated with Eq.(5), assuming $\kappa_{pm} = \kappa_{mp}^* = \kappa \in \text{Re}$. For negligible coupling, the resonant frequencies can cross. When the coupling coefficient fulfills condition (ii), an EP appears in the dispersion. Finally, increasing the coupling beyond this point results in an avoided crossing of the resonant frequencies, and marks the onset of the strong coupling regime.

The former theoretical discussion gives no hints as whether the EP can be reached in practice in a real physical system. In the following, taking as an example the electric and magnetic dipole modes, (ED and MD modes), we rigorously show that EPs can be supported by a single dielectric nanoparticle. As depicted in Figure 1c, the detuning can be controlled by varying height or radius, while the coupling can be increased with the conicity. Hence, we introduce a design strategy that enables full control over the detuning and coupling strength between multipolar modes, providing clear guidelines for the realization of EPs at the nanoscale.

We start by investigating the lowest order multipolar modes of a dielectric nanocylinder. As shown in Figure 2a, the nanoparticle has full rotational symmetry along its principal axis (denoted by $C_\infty$), and hosts a horizontal mirror plane perpendicular to it, which we label as $\sigma_z$. Note that, in consequence, the electromagnetic modes supported by the cylinder can be divided into even or odd with respect to $\sigma_z$. Namely, if we denote by $\hat{\sigma}_z$ an operator that reflects the electromagnetic field along $\sigma_z$, its action over a mode of the cylinder is $\hat{\sigma}_z \mathbf{E}_\alpha = \pm \mathbf{E}_\alpha$, depending on whether the mode is even or odd. A pictorial representation of the ED mode and the MD mode is shown in Figure 2b. It can be clearly seen that the first is even with respect to $\sigma_z$, while the second is odd.

As shown in Figure 3d, the resonant frequencies of the ED and MD modes[38,39] can be overlapped by varying the radius. In a simplified view, this can be understood by interpreting the ED mode as a standing wave formed between the lateral walls of the cylinder, and the MD mode as a 'Fabry-Perot' mode that develops between the top and bottom walls of the cavity[23]. An increase in the separation between the lateral walls is then expected to strongly shift the ED resonance, but affect less the MD resonance, so that both can be brought together (see for instance the right panel in Figure 1a, or Figure 3a). Thus, condition (i) can be easily fulfilled for the two modes.

Unfortunately, condition (ii) cannot be realized, since the ED and MD modes cannot couple. They have different sign under $\hat{\sigma}_z$, so that:

$$\kappa_{md} \propto \langle \mathbf{E}_m, \mathbf{E}_d \rangle = \langle \mathbf{E}_m \, \hat{\sigma}_z^\dagger \hat{\sigma}_z \mathbf{E}_d \rangle = -\langle \mathbf{E}_m, \mathbf{E}_d \rangle = 0 \,. \tag{6}$$

In Eq.(6), we have used the fact that $\hat{\sigma}_z$ is unitary, and the perturbation $\Delta V = \Delta R$ can be taken out of the integral; it is just a constant equal to the difference between the radius of the original cylinder and the perturbed one. Importantly, Eq. (6) explains why EPs have not been yet implemented in all-dielectric nanophotonics: all even and odd modes of a nanodisk excited by a normally incident plane wave follow the same rule, and therefore cannot be brought to coalesce.

The only way of coupling the ED and MD modes is by breaking the mirror symmetry in some fashion. For example, as shown in Figure 2c, the cylinder can be transformed into a truncated cone. This asymmetric perturbation drastically alters the field distributions of the original modes. Its effect can be graphically visualized by decomposing the perturbed field into the sum of the original and the perturbation (Figure 2c). Remarkably, the new modes are of mixed electric and magnetic nature, enabling them to mutually couple.

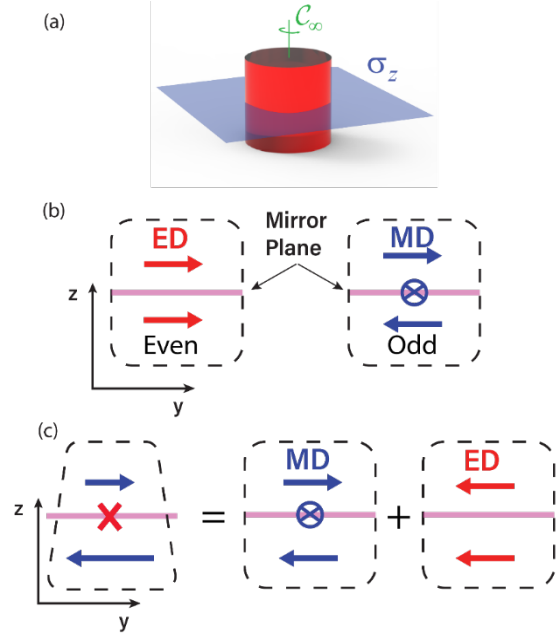

**Figure 2. Coupling of magnetic and electric dipole modes via symmetry breaking.** (a) Symmetry operations defining a cylinder of revolution, where $\mathcal{C}_\infty \parallel \mathbf{z}$ indicates the principal axis of rotation, and $\sigma_z$ a mirror plane perpendicular to the latter. (b) Symmetry of the ED and MD modes with respect to $\sigma_z$. (c) Effect of an out-of-plane perturbation on the MD mode, and equivalent decomposition into a sum of ED and MD contributions.

Having identified the key symmetry that needs to be broken gives us freedom to control at will the parameters entering in Eq.(5). The detuning can be set with the bottom radius $R$, while as we confirm in Figure 3, coupling can be induced by modifying the top-to-bottom ratio $r/R$.

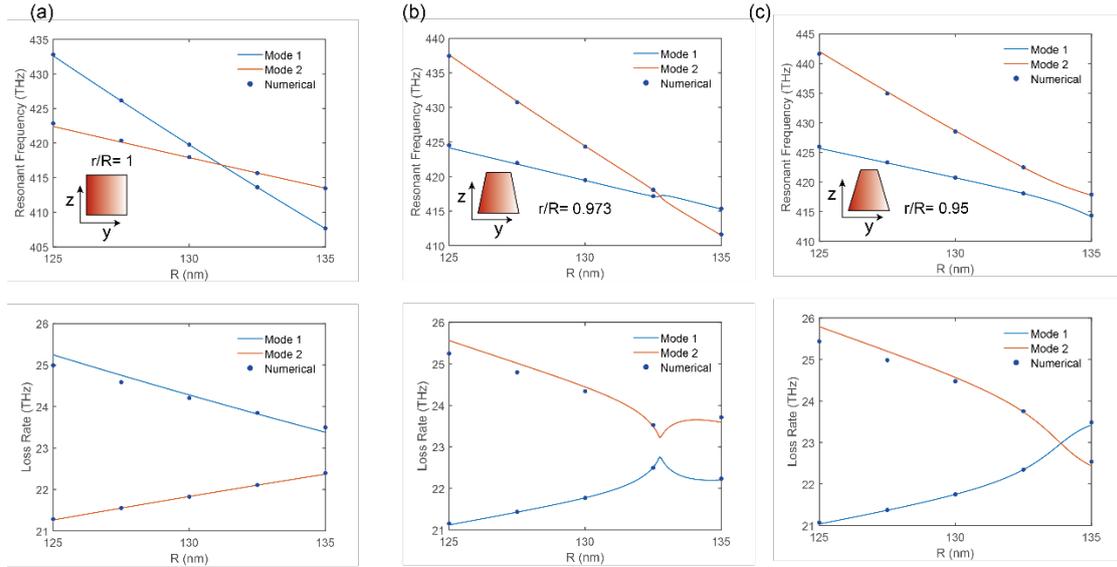

**Figure 3. Comparison between numerical results and the solution of Eqs.(1) for different coupling regimes.** No fitting has been performed. In all cases, the height of the nanoparticle is fixed at $H = 100\,\text{nm}$. (a) Upper panel: crossing of the real parts of the two resonant modes in the weak coupling regime, before transitioning through an EP (r/R=1). The scheme depicts a cross section of unperturbed geometry. Lower panel: avoided crossing of the imaginary parts. (b) same as (a) for r/R=0.95, near the EP. (c) Avoided crossing of the real parts in the strong coupling regime, after transitioning through the EP (upper panel), and crossing of the imaginary parts (lower panel). The schemes in (b)-(c) depict cross sections of the truncated cone (over-deformed for a better visualization). In all plots, lines of different colors indicate different QNMs obtained with Eq.(1). Only when r/R=1 we can associate to each of them a pure electric or magnetic character.

In the next step, we calculate with the help of Eqs. (1) (which do not introduce any assumptions on the coefficients), the evolution of the eigenfrequencies as a function of the two parameters. In this case, $\Delta V(z) = z\Delta r / H + \Delta R$, where the origin of the z-coordinate is taken at the bottom of the nanoparticle, $H$ is the height of the unperturbed cylinder and $\Delta r = r - R$. Note that $\Delta V(z)$ allows investigating both radial deformations and variations in conicity. To demonstrate the accuracy of the analytical method, Figure 3 compares the results of the latter with numerical simulations in COMSOL Multiphysics. Good agreement can be clearly appreciated in both real and imaginary parts for the weak, EP and strong coupling regimes.

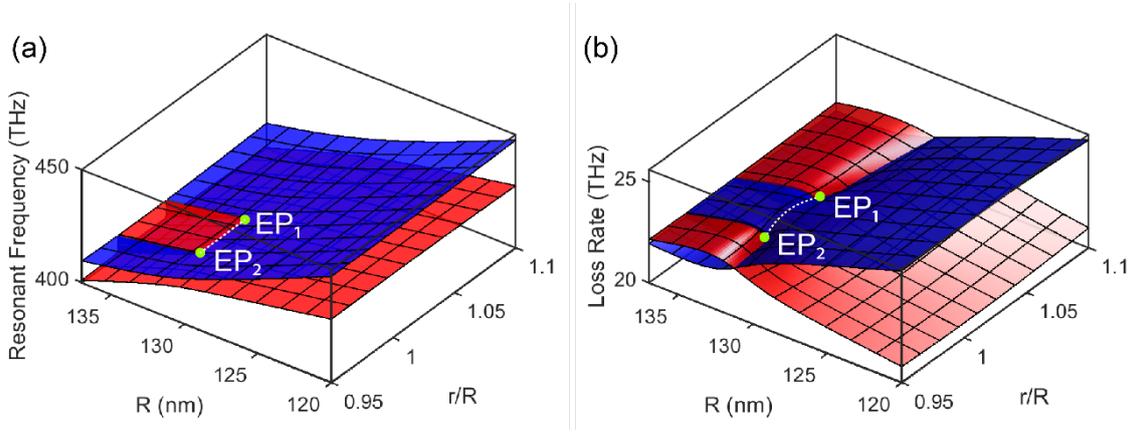

**Figure 4. Calculated Riemann surfaces of the two eigenfrequencies in the vicinity of a pair of EPs.** To obtain them, two parameters of the structure need to be tuned: (i) the radius of a Si cylinder in the vicinity of the Kerker condition (where the ED and MD modes overlap), (ii) the conicity of the particle (ratio between top and bottom radii). (a) Real part, (b) imaginary part (loss rate). White dashed lines indicate the region in parameter space displaying a Fermi arc. Green dots correspond to the EPs, where both real and imaginary parts coalesce.

Having confirmed the validity of our model, we calculate with the latter the evolution of resonant frequencies and loss rates of the two modes as functions of the two parameters, (Figure 4). Starting with a Si nanocylinder with height 100 nm, increasing or decreasing $r/R$ leads to the transition from a crossing of the resonant frequencies to an avoided crossing, a signature of strong coupling. The opposite occurs for the loss rates. Marking the transition from one regime to another, second order EPs can be found (green dots in Figure 3a-b). The two 'eigenfrequency sheets', more commonly known as Riemann surfaces, are seen to intersect each other. The Riemann surface for each mode is 'cut' at an EP, giving rise to rich topological effects. Importantly, for a constant $\Delta R$, $\Delta V(z)$ can take different sign depending on whether the top radius $r$ is increased or decreased. The two situations correspond exactly to the two possible ways to fulfill condition (ii). Thus, EPs in a single dielectric nanoparticle *always come in pairs*. They are connected in parameter space by an open arc, known as a bulk Fermi arc, along which the resonant frequencies of the two modes are degenerate, but display different loss rates. Figure 5 shows a projection of the Riemann surfaces as a function of $r/R$, near (blue lines), and far (orange lines) from the EP pair. After the EPs, there is an abrupt transition and the resonant frequencies split with a dependence $f_0 \propto \sqrt{r/R}$. Drastic changes in the loss rates can also be clearly appreciated. We remark an apparent asymmetry between the conditions to reach EPs for $r/R>1$ and $r/R<1$. This is simply because the Fermi arc is not at constant R, but blueshifts slightly for decreasing $r/R$ due to the change in the overall volume of the nanoparticle.

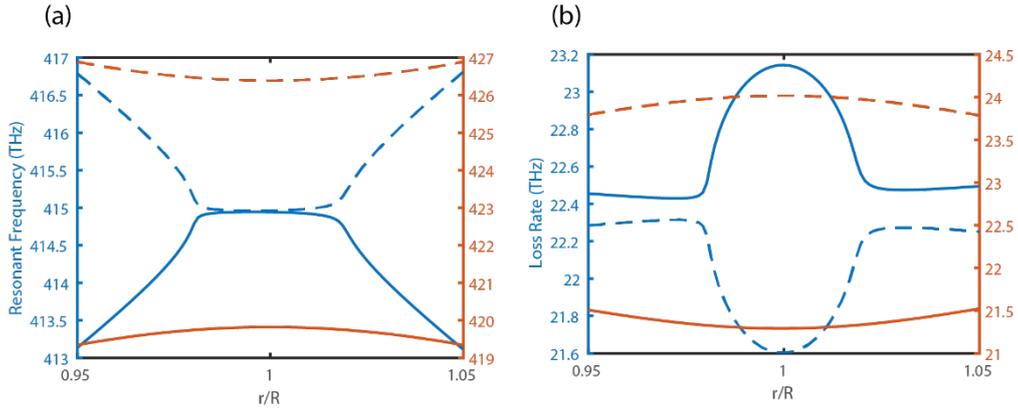

**Figure 5. Projections of the Riemann sheets** in the vicinity of the Fermi arc (blue lines), and far from it (orange lines), calculated for two different bottom radii. Dashed and solid lines indicate the path traced by each of the two QNMs. Near the Fermi arc, the two resonant frequencies are identical, until a critical value of conicity (the EP), where a drastic change in the behavior can be observed.

**Magnetoelectric response in the regime of strong coupling**

Most relevant physical phenomena associated with EPs do not occur exactly at the singularity but in the vicinity of it. For instance, a loop around an EP in parameter space leads to mode interchange. Two loops lead to a recovery of the original mode and an accumulated $4\pi$ 'non-Hermitian' geometrical phase[4]. Other interesting phenomena are associated to the transition from the weak to the strong coupling regime[40,41]. In this section, we demonstrate how, after transitioning through a magnetoelectric EP, the new hybrid modes naturally gain a strong magnetoelectric character. This leads to a bianisotropic response of the nanoparticle.

To do so, we calculate the multipolar decomposition in the weak and strong coupling regimes (Figure 6). Since the perturbation of the cylindrical shape breaks the mirror symmetry, in the strong coupling regime, we observe the emergence of enhanced 'omega-type' bianisotropy[42].

As shown in Figure 6a-b, illumination from the top of the nanoparticle or from bottom results in a drastically different multipolar response, albeit preserving the same total scattering cross section. For comparison, the multipolar decomposition of the unperturbed cylinder is shown in Figure 6c. Interestingly, with illumination from the top, two MD peaks can be clearly seen. However, the ED cross section resembles qualitatively the unperturbed nanocylinder (inset of Figure 6a). Conversely, when the nanocone is illuminated from below, the multipolar spectrum displays the reversed behavior, namely, the ED cross section splits up, while only a single MD peak can be observed.

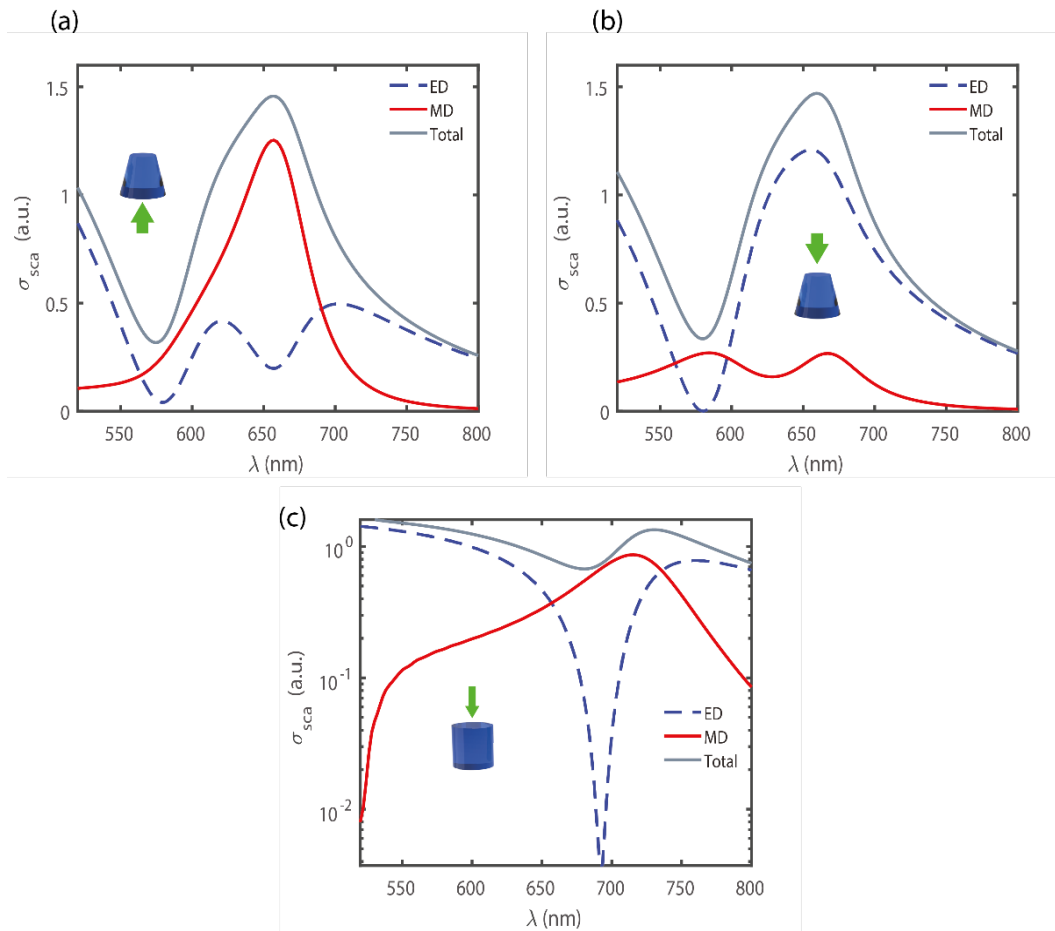

**Figure 6. Bianisotropy and strong coupling**. (a) Multipolar decomposition of the scattering cross section for a truncated cone with R=130 nm, r/R=0.6, H=100 nm, under incident plane wave illumination from the bottom, as depicted in the scheme. (b) Multipolar decomposition of the scattering cross section for the same truncated cone as in (a), with illumination from top. (c) Multipolar decomposition of an unperturbed nanocylinder with R=130 nm.

In Figure 7a, we clarify the origin of the split in the magnetic dipole cross section. The latter is associated with the repulsion of the modes in the strong coupling regime. To verify this, the magnetic dipole cross section is plotted as a function of the bottom radius R for r/R=1 and r/R=0.6 (Figure 7). In the strong coupling regime, the two peaks follow the shift in the resonant frequencies of the hybrid modes. Conversely, in the weak coupling regime (nanocylinder), only one peak is appreciable, corresponding to the MD mode.

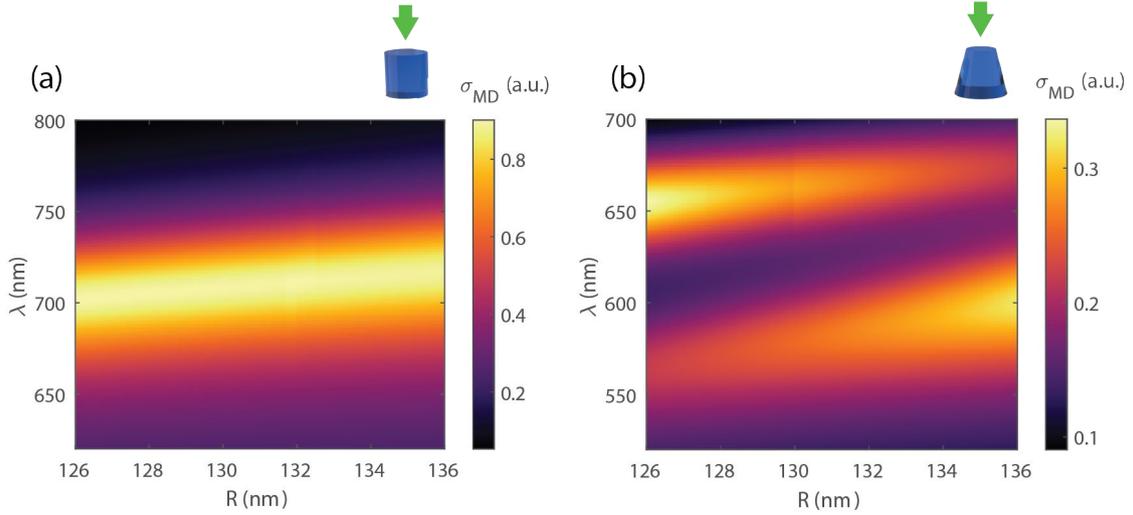

**Figure 7. Multipole decomposition before and after the transition through an EP**. (a)-(b) Magnetic dipole scattering cross section before (a) and after (b) the transition through the EP, for different radii (ratio r/R=0.6). In the second case, a clear split in the MD resonance can be appreciated in the strong coupling regime, in contrast to the weak coupling regime.

**Substrate-mediated control of exceptional points**

Now, we investigate the conditions necessary to design a magnetoelectric EP in a practical experimental setup. As shown in Figure 4a, the latter occurs at very small perturbations of the cylindrical shape (less than 5% change in the radial ratio). This can pose impediments for fabrication. A way to overcome this issue is acting further on the parameters entering condition (ii). Until now, only $\kappa \propto \Delta r / R$ was used. Since the difference between the loss rates of the electric and magnetic dipole modes in the cylinder is originally small, a small $\kappa$ (and therefore a small perturbation of the top radius) is required to compensate it and reach the EP. We propose to introduce a substrate as an additional tuning mechanism. As we explain in the next paragraph, the latter can modify the difference between the loss rates $\Delta\gamma$ (Figure 8a). By doing so, the deformation $\Delta r$ to achieve the EP can be increased to experimentally accessible values.

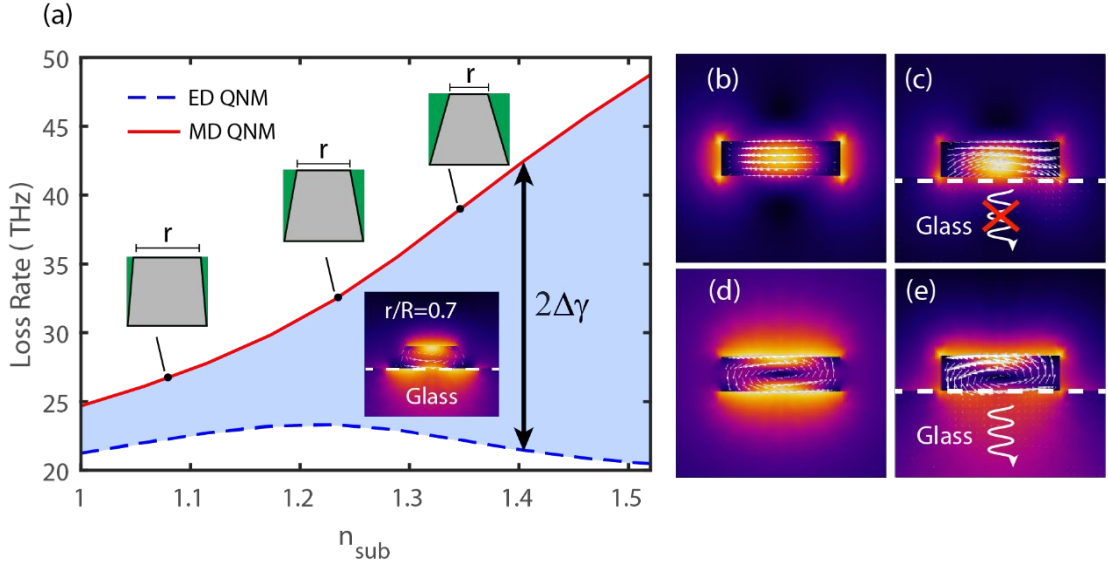

**Figure 8. Substrate engineering of exceptional points.** By controlling the substrate index, the difference between the loss rates of the two involved QNMs can be tuned, thus modifying the critical ratio r/R where the EP will take place. (a) Loss rates of the two QNMs as a function of the substrate index. (b)-(c) Field distribution of the ED QNM in an homogeneous environment (air) (b) and on top of a glass substrate (c). (d)-(e) Same as (c)-(d), but for the MD QNM.

The underlying physical mechanism can be understood once again regarding the ED and MD modes as standing waves of different geometrical origin[23]. As mentioned earlier, the ED mode appears from oscillations between the lateral walls (Mie mode), while the MD mode can be associated with oscillations between the top and bottom walls of the nanocylinder (Fabry-Perot mode). It is thus expected that a modification in the index contrast of the bottom wall of the cavity affects more strongly the MD mode. The eigenfrequency of a Fabry-Perot mode is given, approximately, by[23] $\omega_\ell = c / n_p H \left[ \pi \ell + i \ln\left( r_{top} r_{bott} \right)/2 \right]$, where $\ell$ is the number of maxima of the standing wave, $n_p$ is the particle refractive index, and $r_{top}$, $r_{bott}$ are the reflection coefficients from the top and bottom walls, respectively. The latter expression is derived assuming no coupling with other modes takes place due to the substrate. For high-index nanoparticles deposited on top of dielectric substrates, the assumption is well fulfilled, since the modal fields are strongly confined within the high-index region. From the latter expression, we see that a change in the substrate index affects only the imaginary part of the eigenfrequency, i.e. the loss rate of a Fabry Perot mode evolves approximately as $\gamma \propto \ln(r_{top}r_{bott})$. Conversely, the loss rates of Mie modes are not strongly affected.

The qualitative arguments given above can be exploited as design rules to control the difference between the loss rates of both modes entering condition (ii). To verify our theory, we calculate the loss rates of the ED and MD modes in the nanocylinder for different substrate index contrasts (Figure 8a). We confirm that the ED mode maintains an approximately constant loss rate, while the MD mode almost doubles it when varying the substrate from pure water to glass ($n_g = 1.52$). Figures 8b-e display the field distributions of the ED and MD modes of the nanocylinder in air environment (Figures 8b,d), and when the resonator is placed on top of a glass substrate (Figures 8c,e). In agreement with our physical picture, the ED mode shows no significant changes besides an overall redistribution of the near field. In contrast, the MD mode leaks strongly to the substrate, resulting in an increase of its loss rate.

As a result of the above, the conicity required to reach the magnetoelectric EP increases, as depicted in Figure 8a. On top of glass, the EP occurs near $r/R = 0.7$, an almost 30 % change with respect to the nanoparticle in an homogeneous environment.

Summarizing, we have shown that a dielectric substrate can be used to control the loss rates of Fabry-Perot modes, enabling fine tuning of the EP condition. Specifically, with an increasing substrate index, the EP condition is fulfilled at larger conicities. This mechanism can be exploited to design EPs in nanoparticles with experimentally accessible geometries.

**Design of local refractive-index sensors**

EPs in microcavities have attracted a great deal of interest for the development of ultrasensitive sensor technology[1,11]. Specifically, they can improve significantly the performance of sensing schemes based on the detection of resonance splittings. The basic idea is the following; in a system where two or more modes are degenerate (the sensor), a perturbation $\epsilon$ (the target signal) lifts the degeneracy resulting in a splitting of the resonant frequencies, which can be detected and quantified. For conventional degeneracies (that is, systems where two resonances are simply brought together), the split is linear with the perturbation. However, owing to the topology of the Riemann sheets surrounding the EP, the separation between the resonant frequencies turns out to be in the order of $\sqrt{\epsilon}$. For a sufficiently small perturbation, the sensitivity of an EP is larger than that of a conventional degeneracy[1,9].

In this section, we study the benefits that EPs supported by single nanoparticles could offer for all-optical nanosensors with miniature footprints. Specifically, we envision the possibility to detect local changes in the refractive index of a surrounding environment.

To assess the potential of magnetoelectric EPs in biosensing, we first optimize our truncated nanocone placed on top of a glass substrate and embedded in water (refer to scheme in Figure 9a).

Emulating a conventional experimental protocol[43], we introduce small variations in the refractive index of the surrounding environment ($n_{env}$), ranging from pure water to a concentrated solution of ethylene glycol 9:1 ($n_{env}=1.34$), as shown in Figure 9a. The change in $n_{env}$ introduces a small perturbation to the system's modes, making them depart from the EP. We envision a setup capable of tracking the evolution of the split between the resonant frequencies, $\Delta f$. As a comparison, in Figure 9a we also plot the evolution of $\Delta f$ in a conventional DP formed by overlapping the resonant frequencies of the MD and ED modes of the original nanocylinder. The latter shows the expected linear dependence with $n_{env}$, originated by a distortion of the evanescent fields of the subwavelength modes. In stark contrast, near the EP, $\Delta f$ demonstrates a square-root behavior. To further confirm the power laws, Figure 9b displays a log-log plot of $\Delta f$ as a function of the change in the environment refractive index, $\Delta n_{env}$. The dashed lines correspond to a linear fit, with slope m. The DP displays a slope of 1, whereas the EP is well fitted with m=0.5.

Importantly, due to the square-root asymptotic for small $\Delta n_{env}$, frequency splits near the EP are much larger than those of a conventional dielectric disk optimized at a DP. In consequence, we anticipate optical biosensing schemes based on magnetoelectric EPs to be more sensitive to local environment changes than previously reported devices implemented with conventional ED and MD resonances.

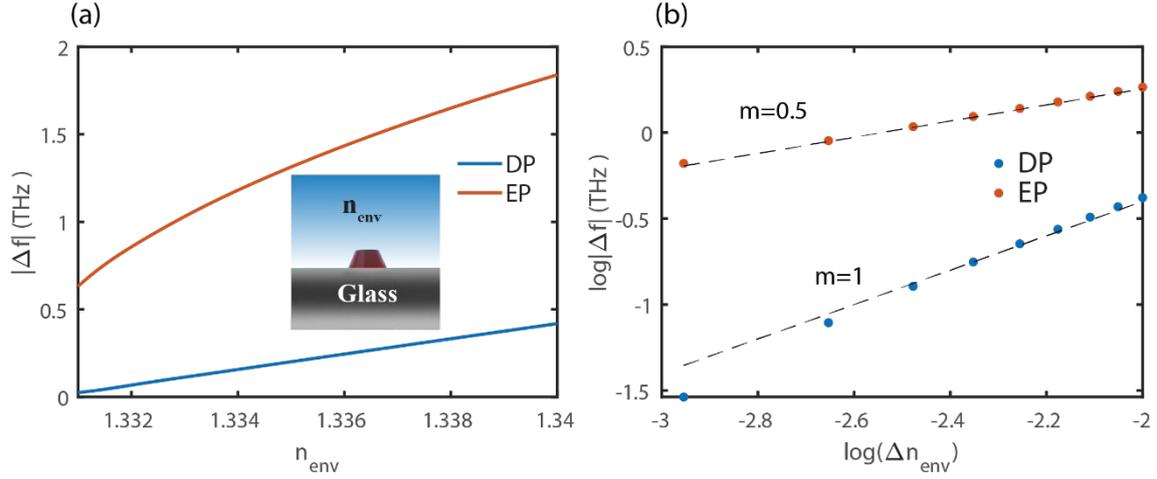

**Figure 9. Refractive index sensitivity near a magnetoelectric exceptional point.** The nanocone dimensions are R=191nm, r/R=0.575, H=100 nm. (a) Split between the resonance frequencies of the two modes when collapsed in an EP or simply brought together by tuning the radius of the nanocylinder, forming a Diabolic Point (DP). The local refractive index is varied from pure water to a solution of ethylene glycol in water ($n_{env} = 1.34$), emulating realistic experimental conditions [43]. (b) Log-Log plot of (a) and linear fit (dashed lines). The slopes (m) show a linear dependence of the modes at the DP with $\Delta n_{env}$, while a square-root dependence (m=0.5), is confirmed for the EP.

**Conclusion and Outlook**

We have predicted the existence of non-Hermitian singularities, EPs, in a single dielectric nanoparticle, subject solely to geometrical perturbations. To do so, we have proposed and validated a modal theory which evidences the critical role of vertical symmetry breaking to achieve the EP. The latter is general and tells us that any two resonant modes of a dielectric disk whose eigenfrequencies cross with the aspect ratio and have opposite sign with respect to a mirror reflection perpendicular to the z-axis can be brought to an EP by a continuous perturbation that breaks the mirror plane of the disk.

Critically, our approach can be implemented for conventional dielectric materials in the near-IR part of the visible spectra, such as Si. As an example, we demonstrate the formation of a 'magnetoelectric EP' between the ED and MD modes by transforming a nanodisk into a truncated nanocone. We confirm a transition from weak to strong coupling mediated by the EP, and omega-type bianisotropy arising in the strong coupling regime. In the latter case, the avoided crossing can be observed in the multipolar decomposition of the scattering cross section. For the first time to our knowledge, we study the role of a dielectric substrate and build a physical picture on its influence on the EP. Interestingly, we suggest theoretically and verify numerically that the latter can be used to bring the EP to experimentally accessible geometrical parameters.

Finally, we propose to utilize magnetoelectric EPs to detect small local changes in the refractive index of an aqueous environment. The idea is based on tracking the split between the resonant frequencies of the two modes. The numerical results demonstrate that much larger sensitivities can be obtained for the EP in comparison with the ED and MD modes of a nanodisk. This is due to the anomalous square-root dispersion characteristic of EPs, in contrast to the conventional linear one. The results suggest a route towards novel biosensing schemes based on single dielectric nanoparticles tuned near an EP.

Thus, this work sets the basis for bringing together all-dielectric nanophotonics and EP physics, paving the way for exciting applications and new uncharted effects. For instance, isolated dielectric nanoparticles supporting EPs can be combined in clusters, which should enable the realization of higher-order EPs[11]. Another intriguing direction is the study of the interplay between EPs and other scattering 'anomalies' commonly encountered in dielectric nanostructures, such as bound states in the continuum[44], lattice resonances or supercavity modes. All these are associated with high quality factors, which could drastically improve the performance of EP-based dielectric biosensors[45].

Beyond sensing, EPs play a crucial role in the emerging field of non-Hermitian topological physics[6]. By revealing the existence of non-Hermitian singularities in realistic Si nanostructures, all-dielectric nanophotonics presents itself as an ideal playground for testing and validating ideas uprooted in the latter.

**Acknowledgements**

The authors thank Andrey Bogdanov for fruitful discussions. The research was supported by Priority 2030 Federal Academic Leadership Program. The authors gratefully acknowledge the financial support from the Ministry of Science and Higher Education of the Russian Federation (Agreement No. № 075-15-2022-1150). The numerical simulations of exceptional points were partially supported by the Russian Science Foundation grant №23-72-00037. V.B. acknowledges the support of the Latvian Council of Science, project: NEO-NATE, No. lzp-2022/1-0553. Y.K. acknowledges a support from the Australian Research Council (grant DP210101292), as well as the International

**References**

1. Park, J.-H. *et al.* Symmetry-breaking-induced plasmonic exceptional points and nanoscale sensing. *Nat. Phys.* **16**, 462–468 (2020).

2. Lai, Y.-H., Lu, Y.-K., Suh, M.-G., Yuan, Z. & Vahala, K. Observation of the exceptional-point-enhanced Sagnac effect. *Nature* **576**, 65–69 (2019).

3. Zhen, B. *et al.* Spawning rings of exceptional points out of Dirac cones. *Nature* **525**, 354–358 (2015).

4. Doppler, J. *et al.* Dynamically encircling an exceptional point for asymmetric mode switching. *Nature* **537**, 76–79 (2016).

5. Zhou, H. *et al.* Observation of bulk Fermi arc and polarization half charge from paired exceptional points. *Science (80-. ).* **359**, 1009–1012 (2018).

6. Parto, M., Liu, Y. G. N., Bahari, B., Khajavikhan, M. & Christodoulides, D. N. Non-Hermitian and topological photonics: Optics at an exceptional point. *Nanophotonics* **10**, 403–423 (2020).

7. Novitsky, D. V., Karabchevsky, A., Lavrinenko, A. V., Shalin, A. S. & Novitsky, A. V. PT symmetry breaking in multilayers with resonant loss and gain locks light propagation direction. *Phys. Rev. B* **98**, 125102 (2018).

8. Yi, C. H., Kullig, J. & Wiersig, J. Pair of Exceptional Points in a Microdisk Cavity under an Extremely Weak Deformation. *Phys. Rev. Lett.* **120**, 93902 (2018).

9. Chen, W., Kaya Özdemir, Ş., Zhao, G., Wiersig, J. & Yang, L. Exceptional points enhance sensing in an optical microcavity. *Nature* **548**, 192–196 (2017).

10. Park, S. H. *et al.* Observation of an exceptional point in a non-Hermitian metasurface. *Nanophotonics* **9**, 1031–1039 (2020).

11. Hodaei, H. *et al.* Enhanced sensitivity at higher-order exceptional points. *Nature* **548**, 187–191 (2017).

12. Xu, H., Mason, D., Jiang, L. & Harris, J. G. E. Topological energy transfer in an optomechanical system with exceptional points. *Nature* **537**, 80–83 (2016).

13. Gao, T. *et al.* Observation of non-Hermitian degeneracies in a chaotic exciton-polariton billiard. *Nature* **526**, 554–558 (2015).

14. Kuznetsov, A. I., Miroshnichenko, A. E., Brongersma, M. L., Kivshar, Y. S. & Luk'yanchuk, B. Optically resonant dielectric nanostructures. *Science (80-. )*. **354**, (2016).

15. Liu, T., Xu, R., Yu, P., Wang, Z. & Takahara, J. Multipole and multimode engineering in Mie resonance-based metastructures. *Nanophotonics* **9**, 1115–1137 (2020).

16. Vestler, D. *et al.* Circular dichroism enhancement in plasmonic nanorod metamaterials. *Opt. Express* **26**, 17841 (2018).

17. Person, S. *et al.* Demonstration of zero optical backscattering from single nanoparticles. *Nano Lett.* **13**, 1806–1809 (2013).

18. Shamkhi, H. K. *et al.* Transverse Scattering and Generalized Kerker Effects in All-Dielectric Mie-Resonant Metaoptics. *Phys. Rev. Lett.* **122**, 193905 (2019).

19. Shamkhi, H. K. *et al.* Transparency and perfect absorption of all-dielectric resonant metasurfaces governed by the transverse Kerker effect. *Phys. Rev. Mater.* **3**, 1–10 (2019).

20. Barhom, H. *et al.* Biological Kerker Effect Boosts Light Collection Efficiency in Plants. *Nano Lett.* **19**, 7062–7071 (2019).

21. Kuznetsov, A. V. *et al.* Special scattering regimes for conical all-dielectric nanoparticles. *Sci. Rep.* **12**, 21904 (2022).

22. Koshelev, K., Favraud, G., Bogdanov, A., Kivshar, Y. & Fratalocchi, A. Nonradiating photonics with resonant dielectric nanostructures. *Nanophotonics* **8**, 725–745 (2019).

23. Canós Valero, A. *et al.* Theory, Observation, and Ultrafast Response of the

Hybrid Anapole Regime in Light Scattering. *Laser Photon. Rev.* 2100114 (2021) doi:10.1002/lpor.202100114.

24. Canós Valero, A. *et al.* Nanovortex-Driven All-Dielectric Optical Diffusion Boosting and Sorting Concept for Lab-on-a-Chip Platforms. *Adv. Sci.* **7**, 1903049 (2020).

25. Murai, S., Castellanos, G. W., Raziman, T. V., Curto, A. G. & Rivas, J. G. Enhanced Light Emission by Magnetic and Electric Resonances in Dielectric Metasurfaces. *Adv. Opt. Mater.* **8**, (2020).

26. Kuznetsov, A. I., Miroshnichenko, A. E., Fu, Y. H., Zhang, J. & Lukyanchukl, B. Magnetic light. *Sci. Rep.* **2**, 1–6 (2012).

27. Rybin, M. V. *et al.* High- Q Supercavity Modes in Subwavelength Dielectric Resonators. *Phys. Rev. Lett.* **119**, 1–5 (2017).

28. Odit, M. *et al.* Observation of Supercavity Modes in Subwavelength Dielectric Resonators. *Adv. Mater.* **33**, 1–7 (2021).

29. Valero, A. C. *et al.* Superscattering Empowered by Bound States in the Continuum. arXiv:2105.13119 (2021) (2021).

30. Bulgakov, E., Pichugin, K. & Sadreev, A. Exceptional points in a dielectric spheroid. *Phys. Rev. A* **104**, 053507 (2021).

31. Terekhov, P. D., Evlyukhin, A. B., Shalin, A. S. & Karabchevsky, A. Polarization-dependent asymmetric light scattering by silicon nanopyramids and their multipoles resonances. *J. Appl. Phys.* **125**, 173108 (2019).

32. Kucherik, A. *et al.* Nano-Antennas Based on Silicon-Gold Nanostructures. *Sci. Rep.* **9**, 338 (2019).

33. Lalanne, P., Yan, W., Vynck, K., Sauvan, C. & Hugonin, J. P. Light Interaction with Photonic and Plasmonic Resonances. *Laser Photonics Rev.* **12**, 1–38 (2018).

34. Defrance, J. & Weiss, T. On the pole expansion of electromagnetic fields. *Opt. Express* **28**, 32363 (2020).

35. Doost, M. B., Langbein, W. & Muljarov, E. A. Resonant-state expansion applied to three-dimensional open optical systems. **013834**, 1–14 (2014).

36. Yan, W., Lalanne, P. & Qiu, M. Shape Deformation of Nanoresonator: A Quasinormal-Mode Perturbation Theory. *Phys. Rev. Lett.* **125**, 013901 (2020).

37. Bogdanov, A. A. *et al.* Bound states in the continuum and Fano resonances in the strong mode coupling regime. *Adv. Photonics* **1**, 1 (2019).

38. Kruk, S. & Kivshar, Y. Functional meta-optics and nanophotonics govern by Mie resonances. *ACS Photonics* **4**, 2638–2649 (2017).

39. Fu, Y. H., Kuznetsov, A. I., Miroshnichenko, A. E., Yu, Y. F. & Luk'yanchuk, B. Directional visible light scattering by silicon nanoparticles. *Nat. Commun.* **4**, 1527 (2013).

40. Djorwe, P., Pennec, Y. & Djafari-Rouhani, B. Frequency locking and controllable chaos through exceptional points in optomechanics. *Phys. Rev. E* **98**, 032201 (2018).

41. Rodriguez, S. R. K. Classical and quantum distinctions between weak and strong coupling. *Eur. J. Phys.* **37**, 25802 (2016).

42. Markovich, D. *et al.* Enhancement of artificial magnetism via resonant bianisotropy. *Sci. Rep.* **6**, 1–8 (2016).

43. Bosio, N. *et al.* Plasmonic versus All-Dielectric Nanoantennas for Refractometric Sensing: A Direct Comparison. *ACS Photonics* **6**, 1556–1564 (2019).

44. Novitsky, D. V. *et al.* CPA-Lasing Associated with the Quasibound States in the Continuum in Asymmetric Non-Hermitian Structures. *ACS Photonics* **9**, 3035–3042 (2022).

45. Wiersig, J. Sensors operating at exceptional points: General theory. *Phys. Rev. A* **93**, 1–9 (2016).